\documentclass[twocolumn]{autart}    
\usepackage[T1]{fontenc}
\usepackage{caption,color}
\usepackage{graphicx}
\usepackage{amsmath,amssymb,amsfonts}
\usepackage{textcomp}
\usepackage{dsfont}
\usepackage{bbold}
\usepackage{arydshln}
\usepackage{comment}
\usepackage{mathtools} 
\usepackage{mathrsfs}
\usepackage{enumerate}
\everymath{\displaystyle}
\usepackage{cases}
\usepackage[framed,thmmarks]{ntheorem}  
\usepackage[overload]{empheq}
\usepackage{booktabs}
\usepackage{datetime}
\usepackage{cite}
\usepackage{stfloats}
\usepackage{pifont}
\newenvironment{proofa}{{\noindent\bfseries{Proof:}}}{\hfill$\blacksquare$}
\usepackage{etoolbox}
\newtheorem{problem}{Problem}
\newtheorem{proposition}{Proposition}
\newtheorem{lemma}{Lemma}
\newtheorem{theorem}{Theorem}
\newtheorem{condition}{Condition}
\newtheorem{remark}{Remark}
\newtheorem{corollary}{Corollary}
\newtheorem*{stassumption}{Standing Assumption}
\AtEndEnvironment{problem}{\hfill$\Box$}
\AtEndEnvironment{lemma}{\hfill$\Box$}
\AtEndEnvironment{theorem}{\hfill$\Box$}
\AtEndEnvironment{assumption}{\hfill$\Box$}
\AtEndEnvironment{condition}{\hfill$\Box$}
\AtEndEnvironment{remark}{\hfill$\Box$}
\AtEndEnvironment{corollary}{\hfill$\Box$}
\AtEndEnvironment{stassumption}{\hfill$\Box$}

\newcommand{\RNum}[1]{\uppercase\expandafter{\romannumeral #1\relax}}

\newcommand{\R}{\mathds{R}}

\newcommand{\skof}{\sum_{k=\tau}^{\infty}}
\newcommand{\ubf}{\mathbf{u}}

\newcommand{\oalpha}{\overline{\alpha}}
\newcommand{\oa}{\overline{a}}

\newcommand{\ua}{\underline{a}}

\newcommand{\ualpha}{\underline{\alpha}}
\newcommand{\sx}{\sigma(x)}

\newcommand{\Zge}{\mathds{Z}_{\geq 0}}

\newcommand{\Zsg}{\mathds{Z}_{\geq 1}}
\newcommand{\Zgeone}{\mathds{Z}_{\geq 1}}

\newcommand{\qdef}{\mathcal{Q}}
\newcommand{\udef}{\mathcal{U}(q)}
\newcommand{\ual}{\ua_{\ell}}
\newcommand{\oav}{\oa_{V}}

\newcommand{\ogamma}{\overline{\gamma}}
\newcommand{\ugamma}{\underline{\gamma}}
\newcommand{\ekl}{\exp-\mathcal{KL}}

\newcommand{\Kinf}{\mathcal{K}_{\infty}}
\newcommand{\ov}{\overline{v}}
\newcommand{\ow}{\overline{w}}
\newcommand{\uw}{w}

\newcommand{\usq}{u^*_0(q)}

\newcommand{\uell}{\underline{\ell}}
\newcommand{\tvartheta}{\overline{\vartheta}}

\begin{document}
\begin{frontmatter}
\title{Stability analysis of
optimal control problems with time-dependent costs}
\thanks[footnoteinfo]{This work was supported by the France Australia collaboration project IRP-ARS CNRS  and the Australian Research Council under the Discovery Project DP210102600.}
\author[nancy]{Sifeddine Benahmed}\ead{sifeddine.benahmed@univ-lorraine.fr},    
\author[nancy]{Romain Postoyan}\ead{romain.postoyan@univ-lorraine.fr},               
\author[melbourne]{Mathieu Granzotto}\ead{mgranzotto@unimelb.edu.au},
\author[romania]{Lucian Bu\c soniu}\ead{lucian.busoniu@aut.utcluj.ro},
\author[nancy]{Jamal Daafouz}\ead{jamal.daafouz@univ-lorraine.fr},
\author[melbourne]{Dragan Ne\v{s}i\'c}\ead{dnesic@unimelb.edu.au}
 \address[nancy]{Universit\'e de Lorraine, CNRS, CRAN, F-54000 Nancy, France}  
\address[melbourne]{Department of Electrical and Electronic Engineering, University of Melbourne, Parkville, VIC 3010, Australia}      
\address[romania]{Department of Automation, Technical University of Cluj-Napoca, Memorandumului 28, 400114 Cluj-Napoca,
Romania}  
\begin{abstract}
We present stability conditions for deterministic time-varying nonlinear discrete-time systems whose inputs aim to minimize an infinite-horizon time-dependent cost. Global asymptotic and exponential stability properties for general attractors are established. This work covers and generalizes the related results on discounted optimal control problems to more general systems and cost functions.
\end{abstract}
\end{frontmatter}

\section{Introduction}
Optimal control deals with the problem of selecting the control inputs such that a cost function is minimized during system operation \cite{Kirk}. Optimal control has a wide range of applications, which go beyond control engineering, including artificial intelligence \cite{sutton2018reinforcement}, energy management \cite{munteanu2008optimal}, economics \cite{KELLETT20197}, medicine \cite{HAYHOE2021495}, and so on. A major question when studying optimal control problems for safety-critical systems is whether the induced closed-loop system exhibits stability  properties. The links between stability and optimality are well-understood in a number of cases including linear systems with quadratic costs \cite{anderson2007optimal}, model predictive control e.g., \cite{RAWLINGS1994185,grimm}, classes of nonlinear systems \cite{sepulchre2012constructive}, to cite a few. Nevertheless, when the cost function depends on time, the question of the stability of the system controlled by optimal inputs remains a challenging and largely open problem. Most results in this case are recent and concentrate on specific time-dependent cost functions, namely discounted costs \cite{Gaitsgory_global_stability_ct, Post_17,Matheo_tac_1}, which are popular in the dynamic programming \cite{Bert05} and reinforcement learning literature \cite{sutton2018reinforcement}, or on classes of time-varying finite-horizon costs \cite{Risbeck}. There is therefore a need for stability conditions applicable to optimal control problems with more general time-dependent costs.

In this context, we investigate the optimal control of deterministic time-varying nonlinear discrete-time systems whose inputs minimize an infinite-horizon time-dependent cost. Our goal is to identify conditions on the plant model and the cost function under which the closed-loop system exhibits stability properties. The challenges when dealing with time-varying systems and time-dependent costs are, first, that the Bellman equation is time-varying and does not admit a convenient time-invariant formulation as for discounted costs, and, second, that the attractor is an unbounded  set, which are both hard to deal with. To address these challenges, we first augment the state of the system with a clock variable, which counts the time, so that the augmented system and the cost become time-invariant with respect to these new coordinates. By doing so, the system dynamics becomes autonomous, which eases the manipulation of Bellman equation, that plays a key role in the stability analysis, while still capturing the features of the problem. Then, we consider stabilizability and detectability conditions on the system and the stage cost, which generalizes the related assumptions in \cite{grimm}, \cite{Post_17} to time-varying systems and costs and which are consistent with the conditions used for linear-quadratic optimal control \cite{anderson2007optimal}.
We rely for this purpose on a generic state measure as in, e.g., \cite{grimm,Post_17,Matheo_tac_1}, which allows covering the situation where the attractor is a closed unbounded set. Based on these assumptions, we provide sufficient conditions involving the optimal value function under which the closed-loop system exhibits a global asymptotic stability property. This property becomes exponential and/or uniform under extra conditions. To facilitate the investigation of the aforementioned sufficient conditions, we present easier-to-check conditions, which do not involve the optimal value function. The derived results cover discounted costs as a special case and are applicable to a much broader type of time-dependent costs.

The rest of the paper is organized as follows. Preliminaries are given in Section~\ref{section:Notation_Preliminaries}. In Section~\ref{section:objective}, we present the general objective and formulate the problem. Section~\ref{section:assumptions} states the detectability and stabilizability conditions. In Section~\ref{Section:stability_guarantees}, we present the main stability results. Section~\ref{Section:ensuring_item_ii_theorem_1} provides sufficient conditions under which the stability results of Section~\ref{Section:stability_guarantees} apply. An illustrative example is given in Section~\ref{Section:illustrative:example}, and Section~\ref{section:conculsion} concludes the paper with some final remarks.

\section{Preliminaries}\label{section:Notation_Preliminaries}
Let $\R$ be the set of real numbers, $\mathds{Z}$ be the set of integers and $\mathds{Z}_{\geq k}:=\{k,k+1,k+2,\ldots\}$ with $k\in\mathds{Z}$. We use $(x,y)$ to denote $[x^{\top},y^{\top}]^{\top}$, where $(x,y)\in\R^n\times\R^m$ and $n,m\in\Zgeone$. The Euclidean norm of a vector $x\in\R^{n}$ is denoted by $|x|$ and the distance of $x\in\R^n$ to a non-empty set $\mathcal{A}\subseteq\R^n$ is denoted by $|x|_{\mathcal{A}}:=\inf\{|x-y| : y \in \mathcal{A}\}$. The notation $\mathds{I}$ stands for the identity map from $\R_{\geq 0}$ to $\R_{\geq 0}$. We consider class-$\mathcal{K}$ and $\mathcal{K}_{\infty}$ functions as defined in  \cite[Chapter 4.4]{Khalil:1173048}.
A continuous function
$\beta:\R_{\geq0}\times\Zge\to\R_{\geq0}$ is of class-$\mathcal{KL}$ when $\beta(s_1,\cdot)$ is decreasing to $0$ for any $s_1\geq0$ and $\beta(\cdot,s_2)$ is of class-$\mathcal{K}$ for any $s_2\in\Zge$.
We write $\beta\in\exp-\mathcal{KL}$ when there exist $\lambda_1\geq1$ and $\lambda_2\in[0,1)$ such that $\beta(s_1,s_2)=\lambda_1\lambda_2^{s_2}s_1$ for any $s_1\geq0$ and $s_2\in\Zge$. Given any  $\alpha:\R_{\geq 0}\times\Zge\to\R_{\geq 0}$, we introduce for the sake of convenience the notation $\alpha_{\tau}$ to denote $\alpha(\cdot,\tau)$ for any $\tau\in\Zge$. Finally, given an infinite-length sequence $\mathbf{z}$, we denote by $\mathbf{z}|_k$ the sequence made of the first $k$ elements of $\mathbf{z}$ where $k\in\Zgeone$.
\section{Objective}\label{section:objective}
\subsection{System and cost function}\label{section:system_and_cost}
Consider the time-varying system 
\begin{align}\label{E:S:intial_system}
    x({k+1})&=f(x(k),u(k),k),
   \end{align}
where $x(k)\in\R^{n_x}$ is the state, $u(k)\in\mathcal{U}(x(k),k)$ is the input at time $k\in\Zge$ and  $\mathcal{U}(x,k)\subseteq \R^{n_u}$ is the non-empty set of admissible inputs associated to state $x$ and time $k$, with $n_x$, $n_u\in\Zgeone$. Given the initial time $\tau\in\Zge$ and initial state $x\in\R^{n_x}$, we denote the solution to \eqref{E:S:intial_system} at time $k\in\mathds{Z}_{\geq\tau}$ with the admissible sequence of inputs $\mathbf{u}:=(u_{\tau},u_{\tau+1},\ldots)$  as $\psi(k,\tau,x,\mathbf{u}|_k)$  with $\psi(\tau,\tau,x,\cdot)=x$, where we recall that by admissible inputs we mean that $u_k\in\mathcal{U}(\psi(k,\tau,x,\mathbf{u}|_k),k)$ for any $k\geq\tau$. 

We investigate the scenario where, for a given initial condition $x\in\R^{n_x}$ and initial time $\tau\in\Zge$, the sequence of admissible inputs applied to \eqref{E:S:intial_system} minimizes the time-dependent cost function
\begin{equation}\label{E:S:intial_cost}
 \mathcal{J}(x,\tau,\mathbf{u}):=\skof\ell(\psi(k,\tau,x,\mathbf{u}|_{k}),k,u_k),
\end{equation}
where $\ell:\R^{n_x}\times\mathds{Z}_{\geq0}\times\R^{n_u}\to \R_{\geq0}$ is the non-negative stage cost. Cost function $\mathcal{J}$ has an infinite horizon and its stage cost depends on the plant state, the control input and the time. Examples of such costs include stage costs made of the product of a state- and input-dependent term with a time-dependent function, thereby covering discounted costs, see Sections~\ref{section:case_ell_is_ell_1_dot_ell_2} and \ref{Section:illustrative:example} for examples.

Assuming there exists at least one sequence of admissible inputs for any initial state $x$ and initial time $\tau$, which minimizes \eqref{E:S:intial_cost} as formalized in the sequel,  our objective is to identify conditions on $f$ and $\ell$ under which  system \eqref{E:S:intial_system} whose inputs minimize \eqref{E:S:intial_cost}  exhibits  stability properties. The problem is formalized next.

\subsection{Problem formulation}\label{section:problem_formulation}
To proceed with the analysis, we augment the state vector $x$ with the clock variable $\tau\in\Zge$, which counts the time, we thus obtain
\begin{subequations}\label{E:S:augmented_system} 
   \begin{numcases}{}
    x(k+1) =f(x(k),u(k),\tau(k)),\\
    \tau(k+1) =\tau(k)+1,\label{subE:S:augmented_system}
    \end{numcases}  
\end{subequations}
where $u(k)\in\mathcal{U}(x(k),\tau(k))$. For the sake of convenience, we write system \eqref{E:S:augmented_system} in the compact form 
\begin{equation}\label{E:S:compact_augmented_system}
    q(k+1)=F(q(k),u(k)),
\end{equation}
with $q(k):=(x(k),\tau(k))\in\qdef$, $\qdef:=\R^{n_x}\times\Zge$, and $F(q,u):=(f(x,u,\tau),\tau+1)$ for any $q\in\qdef$ and $u\in\mathcal{U}(q)$. We denote the solution to \eqref{E:S:compact_augmented_system} at time $k\in\Zge$ initialized at $q=(x,\tau)\in\qdef$ at time $0$ with the admissible sequence of inputs $\mathbf{u}=(u_0,u_{1},\ldots)$ as $\phi(k,q,\mathbf{u}|_k):=(\phi_1(k,q,\mathbf{u}|_k),\phi_2(k,q))$ where $\phi_1(k,q,\mathbf{u}|_k)\in\R^{n_x}$, $\phi_2(k,q)\in\Zge$ and $\phi(0,q,\cdot)=q$; notice that $\phi_2$ does not depend on $\mathbf{u}|_k$. We note that any solution $\psi$ to \eqref{E:S:intial_system} initialized at $x$ at time $\tau$ with admissible sequence of inputs $\ubf$ is equal to the  $\phi_1$-component of the solution to \eqref{E:S:augmented_system} initialized at $(x,\tau)$ at time $0$ with the same sequence of inputs and \emph{vice versa}. For this reason,
we focus on system~\eqref{E:S:compact_augmented_system}, whose solutions are initialized at time 0. We can then write the cost function \eqref{E:S:intial_cost}, for any $q\in\qdef$ and infinite-length admissible sequence of input $\mathbf{u}=(u_0,u_1,\cdots)$, as
\begin{equation}\label{E:S:cost_augmented_system}
    J(q,\mathbf{u}):=\sum_{k=0}^{\infty}\ell(\phi_1(k,q,\mathbf{u}|_k),\phi_2(k,q),u_k).
\end{equation}
 As mentioned at the end of Section~\ref{section:system_and_cost}, we assume that, for any $q\in\qdef$, there exists (at least) one infinite-length sequence of admissible inputs, which minimizes \eqref{E:S:cost_augmented_system}, as formalized below.\\
 
\begin{stassumption}[\textbf{SA}] For any $q\in\qdef$, there exists an infinite-length sequence of admissible inputs $\ubf^{*}(q)$, called optimal solution, such that
    $J(q,\ubf^*(q))=\underset{\ubf}{\min}\ J(q,\ubf) =:V^{\star}(q)$,
    where $V^{\star}$ is the optimal value function.
\end{stassumption}
Conditions on system \eqref{E:S:compact_augmented_system} and cost function \eqref{E:S:cost_augmented_system} to ensure SA can be found in \cite{gilbert}.
SA implies that the set defined below is non-empty for any $q=(x,\tau)\in\qdef$ in view of Bellman equation
\begin{equation}\label{Bellman_equation}
\mathcal{U}^*(q):=\underset{u\in\mathcal{U}(q)}{\mathrm{argmin}}\left\{\ell(q,u)+V^{\star}\big(F(q,u)\big)\right\}.
\end{equation}
Note that, thanks to the above state augmentation, Bellman equation in \eqref{Bellman_equation} is stationary, which is convenient in the sequel to proceed with the analysis. We can then represent system \eqref{E:S:compact_augmented_system} whose sequence of inputs minimizes \eqref{E:S:cost_augmented_system} as the next difference inclusion
{\setlength\arraycolsep{0pt}\begin{equation}\label{E:S:closed_loop_augmented_system}
\begin{array}{rllll}
 q(k+1) \in F^*(q(k)) := \begin{bmatrix}f(x(k),\mathcal{U}^*(q(k)),\tau(k))\\\tau(k)+1\end{bmatrix},
 \end{array}
\end{equation}}
\hspace{-0.15cm}where $f(x,\mathcal{U}^*(q),\tau)=\{f(x,u,\tau) : u \in\mathcal{U}^*(q)\}$ for  $q=(x,\tau)\in\qdef$. We denote the solution to \eqref{E:S:closed_loop_augmented_system} at time $k\in\Zge$ initialized at $q=(x,\tau)\in\qdef$ at time $0$ as $\phi^*(k,q)=:(\phi_1^*(k,q),\phi_2^*(k,q))$ where $\phi_1^*(k,q)\in\R^{n_x}$ and $\phi_2^*(k,q)\in\Zge$.

We are ready to formalize the problem. \\

\begin{problem}\label{problem_1}
Provide conditions on $f$ and $\ell$ under which there exist  $\sigma:\R^{n_x}\to\R_{\geq0}$ continuous and $\beta:\R_{\geq0}\times\Zge\times\Zge\to\R_{\geq0}$ continuous with  $\beta(\cdot,\cdot,s)\in\mathcal{KL}$ for any $s\in\Zge$, such that for any $q=(x,\tau)\in\qdef$, any solution $\phi^*$ to system~\eqref{E:S:closed_loop_augmented_system} initialized at $q$  verifies
\begin{equation}\label{equation:theorem:KL_stability:item_2}
    \sigma(\phi_1^*(k,q))\leq\beta(\sx,k,\tau),
\end{equation}
for any $k\in\Zge$. 
\end{problem}
 The generic function $\sigma$ in Problem~\ref{problem_1} serves as a state measure when investigating stability like in, e.g., \cite{grimm,Post_17, Matheo_tac_1}. When $\sx=|x|$ or $\sx=x^{\top}Px$ for any $x\in\R^{n_x}$, with $P$ a real, symmetric and positive definite matrix and \eqref{equation:theorem:KL_stability:item_2} holds, the set  $\{q=(x,\tau) : \sx=0\}=\{q=(x,\tau) : x=0\}$ is globally asymptotically stable. When $\sx=|x|^p_{\mathcal{A}}$ for any $x\in\R^{n_x}$, with non-empty closed-set $\mathcal{A}\subset\R^{n_x}$ and $p\in\Zgeone$, the set $\{q=(x,\tau) : x\in \mathcal{A}\}$ is globally asymptotically stable when \eqref{equation:theorem:KL_stability:item_2} is satisfied. On the other hand, when \eqref{equation:theorem:KL_stability:item_2} holds with $\beta$ independent of $\tau$, i.e. $\beta\in\mathcal{KL}$, then \eqref{equation:theorem:KL_stability:item_2} is a \textit{uniform} global asymptotic stability property. We recall that by solving Problem~\ref{problem_1}, we  guarantee stability properties for system \eqref{E:S:intial_system} in view of the relations between systems \eqref{E:S:intial_system} and \eqref{E:S:compact_augmented_system} mentioned above. Now, to solve Problem~\ref{problem_1}, we need to present conditions on system~\eqref{E:S:closed_loop_augmented_system}, and thus on $f$ and $\ell$.

\section{Detectability and stabilizability conditions}\label{section:assumptions}
First, $f$ and $\ell$ need to satisfy the next condition, which is related to the detectability of the attractor set for system \eqref{E:S:compact_augmented_system} with respect to output $\ell$, as explained below.\\

\begin{condition}\label{Assumption:Detectability}
 There exist a continuous function $\sigma:\R^{n_x}\to\R_{\geq 0}$, a continuous function $W:\R^{n_x}\times\R^{n_u}\to \R_{\geq0}$ and $\uw,\ow:\R_{\geq0}\times \Zge\to\R_{\geq0}$ where $\uw(\cdot,s)\in\Kinf$ and  $\ow(\cdot,s)$ continuous, non-decreasing and zero at zero for any $s\in\Zge$, such that 
\begin{subequations}\label{equation:assumption_1:inequ}
\begin{align}
     &W(q)\leq\ow(\sx,\tau),\label{equation:assumption_1:inequ_1}\\
   W(F(q,u))-&W(q)\leq - \uw(\sx,\tau)+\ell(q,u),\label{equation:assumption_1:inequ_2}
   \end{align}
\end{subequations}
for any $q=(x,\tau)\in\qdef$ and $u\in\mathcal{U}(q)$.
\end{condition}
Property \eqref{equation:assumption_1:inequ} is related to the detectability of system (\ref{E:S:compact_augmented_system}) with output  $\ell$ with respect to $\sigma$ in view of e.g., \cite{grimm,Hoger_CSL_2019}. Therefore, meeting Condition 2 does not necessarily entail that system~\eqref{E:S:compact_augmented_system} must exhibit any stability property. Condition~\ref{Assumption:Detectability} is satisfied for example when  $\sigma(\cdot)=|\cdot|^2$ and $\ell(q,u)=\ell_1(x,u)\ell_2(\tau)$ where $\ell_1(x,u)=x^{\top}Qx+u^{\top}Gu$ and $\ell_2:\Zge\to\R_{>0}$ for any $q=(x,\tau)\in\qdef$ and $u\in\R^{n_u}$ with $Q$ and $G$ real, symmetric, positive definite and semi-definite respectively. Indeed, by taking $\sx=x^{\top}Qx$, we have $W=0$, $\ow=0$ and $\uw(\sx,\tau)=\ell_2(\tau)\sx$ for any $(x,\tau)\in\qdef$. Condition~\ref{Assumption:Detectability} generalizes \cite[SA3]{grimm} and \cite[item (ii) of Assumption 1]{Post_17} to time-varying systems and cost functions.


Next, we present a condition related to the stabilizability of system~\eqref{E:S:compact_augmented_system}.\\

\begin{condition}\label{Assumption:Stabilizability}
There exists $\ov:\R_{\geq 0}\times\Zge\to\R_{\geq 0}$, where $\ov(\cdot,s)\in\mathcal{K}_{\infty}$ for any $s\in\Zge$, such that for any $q=(x,\tau)\in\qdef$, $V^{\star}(q)\leq\ov(\sx,\tau)$, where $\sigma$ comes from Condition~\ref{Assumption:Detectability}.
\end{condition}
Condition~\ref{Assumption:Stabilizability} is a generalization of \cite[Assumption~1]{Post_17} and  \cite[SA4]{grimm} to time-varying systems and time-dependent costs.
Sufficient conditions that ensure the property stated in Condition~\ref{Assumption:Stabilizability} are given in the next lemma.\\

\begin{lemma}\label{Lemma:example_of_ell_1_for_assumption_1} Suppose that there exist  $\lambda > 0$ and $\alpha:\R_{\geq0}\times\Zge\to\R_{\geq0}$ with $\alpha(\cdot,s)\in\Kinf$ for any $s\in\Zge$, such that for any $q=(x,\tau)\in\qdef$ there exists an admissible infinite-length control input sequence $\mathbf{u}(q)=(u_0(q),u_1(q),\ldots)$ verifying,  for any solution $\phi$ to \eqref{E:S:compact_augmented_system} and $k\in\mathds{Z}_{\geq 0}$,
    \begin{equation}
        \ell(\phi_1(k,q,\mathbf{u}(q)|_{k}),\phi_2(k,q),u_k(q))\leq \alpha(\sigma(x),\tau)e^{-\lambda k}.
    \end{equation}
 Then Condition~\ref{Assumption:Stabilizability} holds with $\ov(s_1,s_2)=\tfrac{1}{1- e^{-\lambda}}\alpha(s_1,s_2)$ for any $s_1\geq0$ and $s_2\in\Zge$.
\end{lemma}
\begin{proofa}
Let $q\in\qdef$ and consider an infinite-length admissible sequence of inputs $\mathbf{u}(q)$ as in Lemma~\ref{Lemma:example_of_ell_1_for_assumption_1}. From the conditions of Lemma~\ref{Lemma:example_of_ell_1_for_assumption_1}, for any $N\in\Zgeone$ and any solution $\phi$ to \eqref{E:S:compact_augmented_system}, 
\begin{flalign}\label{equation:inegalite:proof:lemma:exemple:ell_1}
\sum_{k=0}^{N}\ell(&\phi_1(k,q,\mathbf{u}(q)|_{k}),\phi_2(k,q),u_k(q))\leq\\
&\sum_{k=0}^{N}\alpha(\sx,\tau) e^{-\lambda k} \leq\frac{ 1}{1- e^{-\lambda}}\alpha(\sx,\tau).\nonumber
\end{flalign}
The denominator in the last line of \eqref{equation:inegalite:proof:lemma:exemple:ell_1} is strictly positive as $\lambda>0$. Moreover, the inequalities above hold for any $N\in\Zgeone$, and $N\mapsto \Sigma_{k=0}^{N}\ell(\phi_1(k,q,\mathbf{u}(q)|_{k}),\phi_2(k,q),$ $u_k(q))$ is non-decreasing. Therefore, by taking the limit as $N$ tends to $\infty$, we derive that  $J(q,\mathbf{u}(q))\leq \tfrac{1}{1- e^{-\lambda}}\alpha(\sx,\tau)$. As a result, $V^{\star}(q)\leq J(q,\mathbf{u}(q))\leq\tfrac{1}{1- e^{-\lambda}}\alpha(\sx,\tau)$. This implies the satisfaction of Condition~\ref{Assumption:Stabilizability} with $\ov$ given in Lemma~\ref{Lemma:example_of_ell_1_for_assumption_1}.
\end{proofa}

Lemma~\ref{Lemma:example_of_ell_1_for_assumption_1} is a generalization of \cite[Lemma 1]{Post_17} for time-dependent stage costs, and its condition means that, for a given $\tau\in\Zge$, $\ell$ is  (non-uniformly) globally exponentially stabilizable to zero with respect to $\sigma$ for system \eqref{E:S:compact_augmented_system}, see \cite[Definition 2]{grimm}.\\
\begin{remark} Meeting Condition~\eqref{Assumption:Detectability} does not necessarily entail that system~\eqref{E:S:intial_system} must exhibit exponential stabilization properties. To illustrate this, consider $x(k+1)=0$ when $x(k)\leq0$, $x(k+1)=\tfrac{x(k)}{1+x(k)}$ when $x(k)\geq0$, where $x(k)\in\R$ and $k\in\Zge$. For any $x(0)\geq 0$ and $k\in\Zge$, we have $x(k)=\tfrac{x(0)}{1+kx(0)}$, which means $x(\cdot)$ converges non-exponentially to the origin as time grows. Still, we show below that Condition 2 holds. Let $\ell(x,u)=x^2$ and $\sx=|x|$ for any $x\in\R$ and $u\in\R$. Condition~2 is satisfied with $V^{\star}(x)=x^2$ when $x\leq0$, $V^{\star}(x)= x^2+x^2\Sigma_{k=1}^{\infty}\tfrac{1}{(1+kx)^2}$ when $x\geq0$, and $\ov(s)= s^2+s^2\Sigma_{k=1}^{\infty}\tfrac{1}{(1+ks)^2}$ for any $s\geq0$, which defines a class-$\Kinf$ function. \end{remark}



In the following, we exploit Conditions~\ref{Assumption:Detectability} and \ref{Assumption:Stabilizability} to solve Problem~\ref{problem_1}.

\section{Stability properties}\label{Section:stability_guarantees}
To proceed with the stability of system~\eqref{E:S:closed_loop_augmented_system}, we consider a Lyapunov-like function $Y$ given by $Y:=V^{\star}+W$, where $V^{\star}$ and $W$ come from SA and Condition~\ref{Assumption:Detectability}, respectively.  We first derive useful properties of $Y$ along the solutions to \eqref{E:S:closed_loop_augmented_system}. We then exploit this property  to derive a condition under which the stability of system \eqref{E:S:closed_loop_augmented_system} can be established.
\subsection{Lyapunov-like function and its properties}

The next proposition states key properties of $Y$.\\

\begin{proposition}\label{proposition:relation_Vqk_Vq}
Consider system \eqref{E:S:closed_loop_augmented_system} and suppose Conditions~\ref{Assumption:Detectability} and \ref{Assumption:Stabilizability} hold. Then, the following holds for $Y=V^{\star}+W$.
\begin{itemize}
\item[(i)] For any $q=(x,\tau)\in\qdef$, $\ualpha(\sx,\tau)\leq Y(q)\leq\oalpha(\sx,\tau)$ with $\ualpha=\uw$,  $\oalpha=\ov+\ow$ and where $\uw,\ow$ and $\ov$ come from Conditions~\ref{Assumption:Detectability} and \ref{Assumption:Stabilizability}, respectively.
\item[(ii)] For any $q=(x,\tau)\in\qdef$ and $k\in\mathds{Z}_{\geq0}$, any solution $\phi^*$ to system \eqref{E:S:closed_loop_augmented_system} initialized at $q$ at time $0$ verifies
 \begin{equation}
  Y(\phi^*(k,q))\leq \vartheta^{(k)}\big(Y(q),k+\tau\big), 
 \end{equation}
 where, for any $s_1\geq0$ and $k\in\Zgeone$, $\vartheta^{(0)}(s_1,\cdot) :=  s_1$, $\vartheta^{(k)}(s_1,s_2)  :=  \theta\big(\vartheta^{(k-1)}(s_1,s_2-1),s_2-1\big)$ for any $s_2\in\Zgeone$, and 
 $\theta(s_1,s_2)  := s_1 - \ualpha_{s_2}\circ\overline\alpha^{-1}_{s_2}(s_1)$ for any $s_2\in\Zge$. \hfill $\Box$
\end{itemize}
\end{proposition} 

\begin{proofa}
Let $q=(x,\tau)\in\qdef$ and $q^+\in F^*(q)=(f(x,u_0^*(q),\tau),\tau+1)$ where $u^*_0(q)\in\mathcal{U}^*(q)$, which exists by SA. 
 We first show that item (i) of Proposition~\ref{proposition:relation_Vqk_Vq} is satisfied. In view of Conditions~\ref{Assumption:Detectability} and \ref{Assumption:Stabilizability}, we have \begin{equation}\label{equation:proof:proposition_1:New_1}
    Y(q)\leq\ov(\sx,\tau)+\ow(\sx,\tau)=\oalpha(\sx,\tau).
\end{equation}  
On the other hand, by definition of $V^{\star}$, we derive that $V^{\star}(q)\geq \ell(q,\usq)$. Moreover, by Condition~\ref{Assumption:Detectability}, we have $W(q)\geq W(q^+)+\uw(\sigma(x),\tau)-\ell(q,\usq)\geq\uw(\sx,\tau)-\ell(q,\usq)$. Hence $
Y(q)\geq\uw(\sx,\tau)=\ualpha(\sx,\tau)$. \\
We have proved that item (i) of Proposition~\ref{proposition:relation_Vqk_Vq} holds as $\oalpha(\cdot,s), \ualpha(\cdot,s)\in\Kinf$ for any $s\in\Zge$ in view of Conditions~\ref{Assumption:Detectability} and \ref{Assumption:Stabilizability}.

We now show that item (ii) of Proposition~\ref{proposition:relation_Vqk_Vq} is satisfied. From Bellman equation $V^{\star}(q)=\ell(q,u^*_0(q))+V^{\star}(q^+)$, hence
\begin{align}\label{equation:proposition:case_1:new_3}
    V^{\star}(q^+)-V^{\star}(q)=-\ell(q,u^*_0(q)).
\end{align}
Using \eqref{equation:assumption_1:inequ_2}, we obtain $\ell(q,u^*_0(q))\geq\uw(\sx,\tau)+W(q^+)-W(q)$, thus in view of \eqref{equation:proposition:case_1:new_3},
\begin{align}\label{Eqt:proof_1:bellman_0}
    V^{\star}(q^+)-V^{\star}(q)\leq -\uw(\sx,\tau)-W(q^+)+W(q),
\end{align}
from which it follows
\begin{align}\label{Eqt:proof_1:bellman}
    Y(q^+)-Y(q)\leq -\uw(\sx,\tau)=-\uw_{\tau}(\sx).
\end{align}
On the other hand, for any $s\in\Zge$, the inverse of $\oalpha_s$ exists and is of class-$\mathcal{K}_{\infty}$ as $\oalpha_s\in\mathcal{K}_{\infty}$. Hence, since $Y(q)\leq\oalpha(\sx,\tau)$, we obtain $\oalpha_{\tau}^{-1}(Y(q))\leq \sx$. 
We then derive from \eqref{Eqt:proof_1:bellman} 
\begin{align}\label{Eqt:proof_1:bellman_2}
     Y(q^+)\leq Y(q) -\uw_{\tau} \circ \oalpha_{\tau}^{-1}\left(Y(q)\right).
\end{align}
Therefore, by taking $\theta(s_1,s_2)=s_1-\uw_{s_2}\circ\oalpha_{s_2}^{-1}(s_1)$, which is such that\footnote[1]{This is without loss of generality, as,  if it is not the case, we can always upper-bound, for any $s_2\in\Zge$, $\theta(\cdot,s_2)$ by $\mathds{I}-\Tilde{\alpha}(\cdot,s_2)$, which is of class $\mathcal{K}_{\infty}$, for some suitable $\Tilde{\alpha}(\cdot,s_2)\in\mathcal{K}_{\infty}$ \cite[Lemma B.1]{JIANG2001857}.}   $\theta(\cdot,s_2)\in\mathcal{K}_{\infty}$ for any $s_2\in\Zge$, we derive, by induction from \eqref{Eqt:proof_1:bellman_2}, that for any $q\in\qdef$, any solution $\phi^*$ to \eqref{E:S:closed_loop_augmented_system} initialized at $q$ at time $0$ verifies
\begin{equation}\label{Eqt:proof_2:3}
  Y(\phi^*(k,q)) \leq \vartheta^{(k)}\big(Y(q),k+\tau\big),
 \end{equation}
 for any $k\in\mathds{Z}_{\geq 0}$. Therefore, item (ii) of Proposition~\ref{proposition:relation_Vqk_Vq} holds and the proof is complete.
\end{proofa}

Item (i) of Proposition~\ref{proposition:relation_Vqk_Vq} implies that $Y$ is positive definite and
radially unbounded with respect to $\sigma$ for any $\tau\in\Zge$, and item (ii) of Proposition~\ref{proposition:relation_Vqk_Vq} provides a bound on $Y$ along the solutions to system \eqref{E:S:closed_loop_augmented_system}. We use both  properties in the following to conclude global asymptotic stability properties for system \eqref{E:S:closed_loop_augmented_system}.
\subsection{Global asymptotic stability}\label{section:main_result}
We exploit Proposition~\ref{proposition:relation_Vqk_Vq} to derive the stability property stated in Problem~\ref{problem_1}. We rely for this purpose on an extra condition, which involves $Y$ and thus $V^{\star}$. Since $V^{\star}$ is typically unknown, this condition may be difficult to verify, we thus present special cases where it can be more easily investigated in the sequel.\\

\begin{theorem}\label{Theorem:stability}
Consider system \eqref{E:S:closed_loop_augmented_system} and suppose the following holds.
\begin{itemize}
    \item[(i)] Conditions~\ref{Assumption:Detectability} and \ref{Assumption:Stabilizability} are satisfied.
    \item[(ii)] There exists $\beta:\R_{\geq0}\times\Zge\times\Zge\to\R_{\geq0}$ continuous with $\beta(\cdot,\cdot,s)\in\mathcal{KL}$ for any $s\in\Zge$, such that  for any $q=(x,\tau)\in\qdef$ and $k\in\Zge$,
\begin{equation}\label{Eqt:condition_theorem_1}
\ualpha_{k+\tau}^{-1}\Big(\vartheta^{(k)}\big(Y(q),k+\tau\big)\Big)\leq \beta(\sx,k,\tau),
\end{equation}
\end{itemize}
where $Y$, $\ualpha$ and $\vartheta$ come from Proposition~\ref{proposition:Corollarly:existence_of_L_2_case_alpha=alpha_1alpha_2}. Then any solution $\phi^*=(\phi_1^*,\phi_2^*)$ to system \eqref{E:S:closed_loop_augmented_system} initialized at $q$ at time $0$ satisfies \eqref{equation:theorem:KL_stability:item_2}. Moreover, when $\beta$ is in independent of $\tau$ and of class-$\mathcal{KL}$, the stability property in \eqref{equation:theorem:KL_stability:item_2} is  uniform.
\end{theorem}
\begin{proofa}
Let $q\in\qdef$, $k\in\Zge$ and $\phi^*$ be a solution to system \eqref{E:S:closed_loop_augmented_system} initialized at $q$ at time $0$. We have from item (i) of Proposition~\ref{proposition:relation_Vqk_Vq}$,  Y(\phi^*(k,q))\geq\ualpha(\sigma(\phi_1^*(k,q)),\phi_2^*(k,q))$. In view of item (ii) of Proposition~\ref{proposition:relation_Vqk_Vq}, we have
$Y(\phi^*(k,q)) \leq \vartheta^{(k)}\big(Y(q),k+\tau\big)$. Thus
$\ualpha(\sigma(\phi_1^*(k,q)),\phi_2^*(k,q))\leq\vartheta^{(k)}\big(Y(q),\phi_2^*(k,q)\big)$,
since $\phi_2^*(k,q)=k+\tau$, we derive that 
    $\ualpha(\sigma(\phi_1^*(k,q)),k+\tau)\leq\vartheta^{(k)}\big(Y(q),k+\tau\big)$,
from which it follows
  \begin{equation}\arraycolsep=0.1pt\def\arraystretch{1.2}
\begin{array}{ll}\label{eqt:proof_theorem_stab:042}
    \sigma(\phi_1^*(k,q))\leq\ualpha^{-1}_{k+\tau}\Big(\vartheta^{(k)}\big(Y(q),k+\tau\big)\Big).
\end{array}
 \end{equation}
By applying item (ii) of Theorem~\ref{Theorem:stability}, we obtain
    $\sigma(\phi_1^*(k,q))\leq\beta(\sx,k,\tau)$,
which corresponds to \eqref{equation:theorem:KL_stability:item_2}.
\end{proofa}

Theorem~\ref{Theorem:stability} gives conditions under which Problem~\ref{problem_1} is solved. We note that when item (ii) of Theorem~\ref{Theorem:stability} holds with $\beta(\cdot,\cdot,s)\in\ekl$ for any $s\in\Zge$,  \eqref{equation:theorem:KL_stability:item_2} becomes a global exponential stability property.

As mentioned above, item (ii) of Theorem \ref{Theorem:stability} involves $Y$ and thus $V^{\star}$, which is often unknown. We claim that, still, \eqref{Eqt:condition_theorem_1} can be investigated on a case-by-case basis. To justify this claim, we first provide conditions under which item (ii) of Theorem~\ref{Theorem:stability} holds with $\beta$ independent of $\tau$ in \eqref{Eqt:condition_theorem_1} and thus $\beta\in\mathcal{KL}$ thereby ensuring a uniform global asymptotic stability in this case. We also provide stronger conditions under which $\beta\in\ekl$. Then we focus on an alternative scenario where the stage cost $\ell$ can be written as the product of a state- and input-dependent term with a time-dependent function. We provide explicit conditions on this time-dependent term and the functions $\uw$, $\ow$ and $\ov$ in Conditions~\ref{Assumption:Detectability} and \ref{Assumption:Stabilizability} respectively, under which item (ii) of Theorem~\ref{Theorem:stability} is satisfied, thus ensuring the global asymptotic stability property in \eqref{equation:theorem:KL_stability:item_2}. Finally, extra conditions are given under which $\beta(\cdot,\cdot,s)\in\exp-\mathcal{KL}$ in \eqref{equation:theorem:KL_stability:item_2} for any $s\in\Zge$. \\

\begin{remark}
The stability property defined in Problem~\ref{problem_1} becomes regional if Conditions~1 and 2 are satisfied only in a  subset of the state-space \cite{dragan_teel}, whose interior contains $\{x\in\R^{n_x} : \sx = 0\}\times\Zge$. The stability property may become semiglobal when the stage cost depends on tunable parameters, like a discount factor as shown in \cite{Post_17}. The idea there is to adjust the region of attraction by imposing extra conditions on this parameter.
\end{remark}

\section{Ensuring item (ii) of Theorem~\ref{Theorem:stability}}\label{Section:ensuring_item_ii_theorem_1}
In this section, we first provide sufficient conditions under which item (ii) of Theorem~\ref{Theorem:stability} is ensured and a uniform stability property can be guaranteed, in the sense that $\beta$ in \eqref{Eqt:condition_theorem_1} is independent of $\tau$ and  is of class-$\mathcal{KL}$. We then concentrate on the case where for any $q=(x,\tau)\in\qdef$ and $u\in\udef$, $\ell(q,u)=\ell_1(x,u)\ell_2(\tau)$, where $\ell_1:\R^{n_x}\times\R^{n_u}\to\R_{\geq0}$ and $\ell_2:\Zge\to\R_{>0}$.

\subsection{Sufficient condition for uniform stability properties}
The next proposition provides sufficient conditions under
which item (ii) of Theorem 1 is satisfied with $\beta\in\mathcal{KL}$.\\

\begin{proposition}\label{proposition:Corollarly:existence_of_L_2_case_alpha=alpha_1alpha_2} Suppose the following holds.
\begin{itemize}
   \item[(i)] Condition~\ref{Assumption:Detectability} holds and there exists $\ua\in\Kinf$ such that $\uw(s_1,s_2)\geq\ua(s_1)$ for any $s_1\geq0$, $s_2\in\Zge$. 
      \item[(ii)] Condition~\ref{Assumption:Stabilizability} holds and there exists $\oa\in\Kinf$ such that $\ov(s_1,s_2)+\ow(s_1,s_2)\leq\oa(s_1)$ for any $s_1\geq0$, $s_2\in\Zge$.
      \end{itemize}
 Then item (ii) of Theorem~\ref{Theorem:stability} holds with $\beta$ independent of $\tau$ and of class-$\mathcal{KL}$. \hfill $\Box$
\end{proposition}  
\begin{proofa}
Let $q\in\qdef$ and $k\in\Zge$. We first show that there exists a function $\tvartheta\in\mathcal{KL}$ such that for any $q=(x,\tau)\in\qdef$ and $k\in\mathds{Z}_{\geq0}$,  \begin{equation}\label{Eqt:proof_2:4}
  Y(\phi^*(k,q)) \leq \tvartheta\big(Y(q),k\big).
 \end{equation}
By \eqref{Eqt:proof_1:bellman_2}, in view of items (i) and (ii) of Proposition~\ref{proposition:Corollarly:existence_of_L_2_case_alpha=alpha_1alpha_2} and since $\ualpha=\uw$ and $\oalpha=\ov+\ow$,  we have 
\begin{align}\label{Eqt:proof_prop_case_1_1}
     Y(q^+)-Y(q)\leq -\ualpha_{\tau} \circ \oalpha_{\tau}^{-1}\left(Y(q)\right)\leq- \ua\circ\oa^{-1}(Y(q)).
\end{align}
By \eqref{Eqt:proof_prop_case_1_1}, since $\ua,\oa\in\mathcal{K}_{\infty}$ and in view of \cite[Theorem 8]{NESIC199949}, there exists $\overline\vartheta\in\mathcal{KL}$ such  that \eqref{Eqt:proof_2:4} is satisfied.

By item (i) of Proposition~\ref{proposition:relation_Vqk_Vq}, \eqref{Eqt:proof_2:4} yields $\ua(\sigma(\phi_1^*(k,q)))\leq\tvartheta(\oa(\sx),k),$
from which it follows, $
    \sigma(\phi_1^*(k,q))\leq\ua^{-1}\big(\tvartheta(\oa(\sx),k)\big).$ 
Therefore, item (ii) of Theorem~\ref{Theorem:stability} holds with, for any $s_1\geq0$ and $s_2\in\Zge$,
$\beta(s_1,s_2)=
\ua^{-1}\big(\tvartheta(\oa(s_1),s_2)\big)$,
 the proof is complete.
\end{proofa}

Proposition~\ref{proposition:Corollarly:existence_of_L_2_case_alpha=alpha_1alpha_2} provides conditions under which item (ii) of Theorem~\ref{Theorem:stability} is guaranteed.
These conditions represent some uniform bounds on $\omega$, $\ow$ and $\ov$ coming from Conditions~\ref{Assumption:Detectability} and \ref{Assumption:Stabilizability}, from which the uniform global asymptotic stability property in
(8) follows.

The next corollary presents sufficient conditions under which uniform global \textit{exponential} stability is guaranteed.\\

\begin{corollary}\label{corollary:expo:ell_lower_bounded}
Suppose the following holds.
\begin{itemize}
    \item[(i)] Condition~\ref{Assumption:Detectability} holds and there exists $\ual>0$ such that $\ual s_1\leq\uw(s_1,s_2)$ for any $s_1\geq0$, $s_2\in\Zge$. 
      \item[(ii)] Condition~\ref{Assumption:Stabilizability} holds and there exists $\oav>0$ such that $\ov(s_1,s_2)+\ow(s_1,s_2)\leq\oav s_1$ for any $s_1\geq0$, $s_2\in\Zge$.
         \end{itemize}
        Then item (ii) of Theorem~\ref{Theorem:stability} holds with $\beta$ independent of $\tau$ and of class-$\ekl$. \end{corollary}
\begin{proofa}
We first show that $\ual\leq\oav$. Since $\uw=\ualpha$ and $\ow+\ow=\oalpha$, by item (i) of Proposition~\ref{proposition:relation_Vqk_Vq}, we have $\uw\leq\ov+\ow$. Therefore, in view of items (i) and (ii) of Corollary~\ref{corollary:expo:ell_lower_bounded}, $\ual s_1 \leq \oav s_1 $ for any $s_1\geq0$, and thus $\ual\leq\oav$. Therefore, by \eqref{Eqt:proof_prop_case_1_1}, 
\begin{equation}\label{Eqt:proof_corollary_case_1_4_a}
Y(\phi^*(k,q))\leq\Big(1-\frac{\ual}{\oav}\Big)Y(q),
\end{equation}
for any $q\in\qdef$ and $k\in\Zge$. Note that $(1-\tfrac{\ual}{\oav})\in[0,1)$ in \eqref{Eqt:proof_corollary_case_1_4_a}. Hence, by induction and with similar steps as in the proof of Proposition~\ref{proposition:Corollarly:existence_of_L_2_case_alpha=alpha_1alpha_2} we obtain, for any $q\in\qdef$ and $k\in\Zge$,
    $\sigma(\phi_1^*(k,q))\leq\frac{\oav}{\ual}\Big(1-\frac{\ual}{\oav}\Big)^k\sx$.
Therefore, item (ii) of Theorem~\ref{Theorem:stability} holds with, for any $s_1\geq 0$ and $s_2\in\Zge$,
$ \beta(s_1,s_2)=\lambda_1s_1\lambda_2^{s_2},
 $
 where $\lambda_1=\tfrac{\oav}{\ual }$ and $\lambda_2=\big(1-\tfrac{\ual}{\oav}\big)$, which is of class $\ekl$. 
\end{proofa}


\subsection{When $\ell(q,u)=\ell_1(x,u)\ell_2(\tau)$ for any $q=(x,\tau)\in\mathcal{Q}$ and $u\in\mathcal{U}(q)$}\label{section:case_ell_is_ell_1_dot_ell_2}
 In this section, we focus on stage costs $\ell$, which can be written as the product of a state- and input-dependent term with a time-dependent function. This class of stage cost includes various examples as shown in Table~\ref{tab:exemple_ell_2}. 
 \def\arraystretch{0.5}
\begin{table*}[t]
 \begin{center}
    \begin{tabular}{llllll}
        Expression  of $\ell_2(k)$ &         parameters  &            \multicolumn{4}{l}{Item (v) of Proposition~\ref{proposition:cost_gamma}}\\
       & & $\ugamma$ & $\ogamma$ & $c_1$ &$c_2$  \\
       \toprule
     (a) \quad
       $\ell_2(k)\in[a,b]$ &  $0<a\leq b <\infty$  &$1$&$1$ &$a$&$b$ \\
    &&&&&\\
      (b) \quad  $\gamma^{k}$ &  $\gamma\in(1-L,\infty)$ & $\gamma$ &$\gamma$ &$1$&$1$ \\
      (c) \quad  $\frac{1}{k^{h}+1}$ & $h> 0$, $L\in\Big(1-\tfrac{1}{(1+e^h)^2},1\Big)$ &  $\frac{1}{1+e^{h}}$ & $1$&$1$&$1$ \\
        (d) \quad      $e^{-\frac{1}{2}|\frac{k-\mu}{m}|}$& $m > 0$, $\mu\in\Zge$, $L\in(1-e^{-1/m},1) $ & $e^{-\tfrac{1}{2m}}$ &$1$ &$e^{-\tfrac{\mu}{2m}}$ & $1$\\
                \bottomrule
    \end{tabular}
    \caption{Examples of functions$^1$ $\ell_2$ in Section~\ref{section:case_ell_is_ell_1_dot_ell_2}:  (a) $\ell_2$ is uniformly lower- and upper-bounded by positive constants $a$ and $b$ respectively; (b) covers both usual discounted costs when $\gamma\in(1-L,1)$ and $L\in(0,1)$ and reverse-discounted costs when $\gamma>1$; (c) is an alternative to the discounted case, for which the decay rate is eventually slower; (d) is a Gaussian-like (Laplacian) function, which can be used when we want that $\ell_2$ increases until a time step $\mu$ then decreases for any $k\geq \mu$.} 
    \label{tab:exemple_ell_2}
  \end{center}
\end{table*}
The next proposition provides sufficient conditions under which item (ii) of Theorem~\ref{Theorem:stability} holds for the case when $\ell_2$  is upper- and lower-bounded by exponential functions.\\

\begin{proposition}\label{proposition:cost_gamma} Suppose the following holds.
\begin{itemize}
 \item[(i)]   For any $q=(x,\tau)\in\qdef$ and $u\in\udef$, $\ell(q,u)=\ell_1(x,u)\ell_2(\tau)$, where $\ell_1:\R^{n_x}\times\R^{n_u}\to\R_{\geq0}$ and $\ell_2:\Zge\to\R_{>0}$.
 \item[(ii)] Condition~\ref{Assumption:Detectability} holds and there exists $\ua\in\mathcal{K}_{\infty}$ such that $\uw(s_1,s_2)\geq\ua(s_1)\ell_2(s_2)$ for any $s_1\geq0$, $s_2\in\Zge$.
  \item[(iii)] Condition~\ref{Assumption:Stabilizability} holds and there exists $\oa\in\Kinf$ such that $\ov(s_1,s_2)+\ow(s_1,s_2)\leq\oa(s_1)\ell_2(s_2)$ for any $s_1\geq0$, $s_2\in\Zge$. 
     \item[(iv)] There exists $L\in(0,1)$ such that for any $s\geq 0$, $L\oa(s)\leq\ua(s)$. 
    \item[(v)]  There exist $c_1,c_2>0$ and $\ugamma,\ogamma\in(1-L,\infty)$  such that   for any $k\in\Zge$, $c_1\ugamma^k \leq \ell_2(k)\leq c_2 \ogamma^k$.
    \end{itemize}
 Then item (ii) of Theorem~\ref{Theorem:stability} holds with $\beta(\cdot,\cdot,s)\in\mathcal{KL}$ for any $s\in\Zge$. Moreover,  when $\ugamma=\ogamma$, $\beta$ is independent of $\tau$ and of class-$\mathcal{KL}$. \hfill $\Box$
\end{proposition}  
\begin{proofa} 
We first show that, for any $q=(x,\tau)\in\qdef$, $k\in\mathds{Z}_{\geq0}$, 
 \begin{equation}\arraycolsep=0.1pt\def\arraystretch{1.2}\label{Equation:majoration_de_vartheta_(k)}
\begin{array}{ll}
  \vartheta^{(k)}&(Y(q),k+\tau)\leq(1-L)^kY(q),
 \end{array}
 \end{equation}
 where $\vartheta$ is defined in Proposition \ref{proposition:relation_Vqk_Vq}. Let $q=(x,\tau)\in\qdef$, by item (iv) of Proposition~\ref{proposition:cost_gamma} we have
\begin{equation}\label{Equation:proof:proposition_cost_gamma:N_1}
L\oa\circ\oalpha_{\tau}^{-1}(Y(q))\leq\ua\circ\oalpha_{\tau}^{-1}(Y(q)),
\end{equation}
where $\oalpha$ comes \footnotetext[1]{The values of the constants $c_1, c_2,\ugamma$ and $\ogamma$ given in Table~\ref{tab:exemple_ell_2} are non-unique.}from item (i) of Proposition~\ref{proposition:relation_Vqk_Vq}. On the other hand, in view of item (iii) of Proposition~\ref{proposition:cost_gamma}, and since $\ell_2(\Zge)\subset\R_{>0}$, we have 
$\oalpha_{\tau}^{-1}(Y(q))\geq\oa^{-1}\left(\tfrac{Y(q)}{\ell_2(\tau)}\right)$. 
Hence $Y(q)\geq \overline\alpha_\tau\circ\overline a^{-1}\big(\tfrac{Y(q)}{\ell_2(\tau)}\big)$ and \eqref{Equation:proof:proposition_cost_gamma:N_1} yields
\begin{equation}
L\dfrac{Y(q)}{\ell_2(\tau)}\leq\ua\circ\oalpha_{\tau}^{-1}(Y(q)).
\end{equation}
We have from Proposition \ref{proposition:relation_Vqk_Vq},
$\vartheta^{(1)}\big(Y(q),\tau+1\big)=Y(q)-\ualpha_{\tau}\circ\oalpha_{\tau}(Y(q))$,
from which it follows in view of item (ii) of Proposition~\ref{proposition:cost_gamma},
\begin{equation}
\label{pE:inequality_L_useful_for_vartheta_2_2}\vartheta^{(1)}\big(Y(q),\tau+1\big)\leq Y(q)-\ua\circ\oalpha_{\tau}^{-1}(Y(q))\ell_2(\tau),
\end{equation}
by Proposition \ref{proposition:relation_Vqk_Vq} and since $\ell_2(\Zge)\subset\R_{>0}$, \eqref{pE:inequality_L_useful_for_vartheta_2_2} yields
\begin{equation}
\label{pE:inequality_L_useful_for_vartheta_2}\vartheta^{(1)}\big(Y(q),\tau+1\big)\leq(1-L)Y(q).
\end{equation}
Then, \eqref{Equation:majoration_de_vartheta_(k)} is derived by induction from \eqref{pE:inequality_L_useful_for_vartheta_2}. 

Let $q=(x,\tau)\in\qdef$ and $k\in\Zge$. We deduce from \eqref{Equation:majoration_de_vartheta_(k)} that
\begin{flalign}\label{Equation:proof_corol_4_1}
\ualpha_{k+\tau}^{-1}\Big(\vartheta^{(k)}\big(Y(q),k+\tau\big)\Big)\leq  \ualpha_{k+\tau}^{-1}\big((1-L)^kY(q)\big).
\end{flalign}
Also, in view of item (ii) of Proposition~\ref{proposition:cost_gamma}, and since $\ell_2(\Zge)\subset\R_{>0}$, we have for any $s_1\geq0$ and $s_2\in\mathds{Z}_{\geq0}$,
\begin{equation}\label{Equation:proof:proposition:1:inverse_of_ualpha}
\ualpha_{s_2}^{-1}(s_1)\leq\ua^{-1}\left(\dfrac{s_1}{\ell_2(s_2)}\right),
\end{equation}
where $\ualpha$ comes from item (i) of Proposition~\ref{proposition:relation_Vqk_Vq}. Therefore, \eqref{Equation:proof_corol_4_1} implies that
\begin{flalign}\label{Equation:proof_corol_5_1_0}
\ualpha_{k+\tau}^{-1}\Big(\vartheta^{(k)}\big(Y(q),k+\tau\big)\Big)\leq\ua^{-1}\Big( \dfrac{(1-L)^k}{\ell_2(k+\tau)}Y(q)\Big).
\end{flalign}
Since  $\ell_2(k+\tau)\geq c_1 \ugamma^{k+\tau}$ by item (v) of Proposition~\ref{proposition:cost_gamma}, 
\begin{flalign}\label{Equation:proof_corol_5_1_1}
\ualpha_{k+\tau}^{-1}\Big(\vartheta^{(k)}\big(Y(q),k+\tau\big)\Big)\leq\ua^{-1}\Big( \dfrac{(1-L)^k}{c_1\ugamma^{k+\tau}}Y(q)\Big).
\end{flalign}
By item (i) of Proposition~\ref{proposition:relation_Vqk_Vq},  we derive that
\begin{flalign}\label{Equation:proof_corol_5_1}
\ualpha_{k+\tau}^{-1}\Big(\vartheta^{(k)}\big(Y(q)&,k+\tau\big)\Big)\leq \\ &\ua^{-1}\Bigg(\dfrac{1}{c_1\ugamma^{\tau}}\Big(\dfrac{1-L}{\ugamma}\Big)^k\oalpha(\sx,\tau)\Bigg).\nonumber
\end{flalign}
Consequently, in view of items (iii) and (v) of Proposition~\ref{proposition:cost_gamma},
\begin{flalign}\label{Equation:proof_corol_5_1_2}
\ualpha_{k+\tau}^{-1}\Big(\vartheta^{(k)}\big(Y(q)&,k+\tau\big)\Big)\leq \\ &\ua^{-1}\bigg(\dfrac{c_2\ogamma^\tau}{c_1\ugamma^{\tau}}\bigg(\dfrac{1-L}{\ugamma}\bigg)^k\oa(\sx)\bigg).\nonumber
\end{flalign}
Finally, item (ii) of Theorem~\ref{Theorem:stability} holds with
    $\beta(s_1,s_2,s_3)=\ua^{-1}\bigg(\tfrac{c_2\ogamma^{s_3}}{c_1\ugamma^{s_3}}\bigg(\tfrac{1-L}{\ugamma}\bigg)^{s_2}\oa(s_1)\bigg)$, for any $s_1\geq0$ and $s_2,s_3\in\Zge$,
which is of class-$\mathcal{KL}$ for any $s_3\in\Zge$, as $\ugamma\in(1-L,\infty)$ and $\ua,\oa\in\mathcal{K}_{\infty}$.  This concludes the proof.
\end{proofa}

Proposition~\ref{proposition:cost_gamma} states that when $\ell_2$ is lower-bounded and upper-bounded respectively by exponential functions $k\mapsto c_1\ugamma^k$ and $k\mapsto c_2\ogamma^k$, under conditions on $\uw$, $\ow$ and $\ov$ coming from Conditions~\ref{Assumption:Detectability} and \ref{Assumption:Stabilizability}, global asymptotic stability is guaranteed.
It is worthy to note that Proposition~\ref{proposition:cost_gamma} includes the case of discounted and reverse-discounted costs when $\ell_2=\ugamma^k=\ogamma^k$ and $c_1=c_2=1$. Again, other examples are given in Table~\ref{tab:exemple_ell_2}.

The next corollary provides sufficient conditions under which  global exponential stability is ensured when $\ell_2$ satisfies item (v) of Proposition~\ref{proposition:cost_gamma}.\\

\begin{corollary}\label{corollary:expo:ell_discount}
Suppose the following holds.
\begin{itemize}
 \item[(i)]   For any $q=(x,\tau)\in\qdef$ and $u\in\udef$, $\ell(q,u)=\ell_1(x,u)\ell_2(\tau)$, where $\ell_1:\R^{n_x}\times\R^{n_u}\to\R_{\geq0}$ and $\ell_2:\Zge\to\R_{>0}$.
 \item[(ii)] Condition~\ref{Assumption:Detectability} holds and there exists $\ual>0$ such that $\ual s_1\ell_2(s_2)\leq\uw(s_1,s_2)$ for any $s_1\geq0$, $s_2\in\Zge$.
  \item[(iii)] Condition~\ref{Assumption:Stabilizability} holds and there exists $\oav>0$ such that $\ov(s_1,s_2)+\ow(s_1,s_2)\leq\oav s_1\ell_2(s_2)$ for any $s_1\geq0$, $s_2\in\Zge$. 
       \item[(iv)]  There exist $c_1,c_2>0$ and $\ugamma,\ogamma\in(1-\tfrac{\ual}{\oav},\infty)$  such that for any $k\in\Zge$, $c_1\ugamma^k \leq \ell_2(k)\leq c_2 \ogamma^k$.
    \end{itemize}
      Then item (ii) of Theorem~\ref{Theorem:stability} holds with $\beta(\cdot,\cdot,s)\in\ekl$ for any $s\in\Zge$. \end{corollary}
      \begin{proofa} We first show that, for any $q=(x,\tau)\in\qdef$ and $k\in\mathds{Z}_{\geq0}$,   \begin{equation}\arraycolsep=0.1pt\def\arraystretch{1.2}\label{Equation_0:proof:corollary_3}
\begin{array}{ll}
  \vartheta^{(k)}&(Y(q),k+\tau)\leq\Big(1-\dfrac{\ual}{\oav}\Big)^kY(q),
 \end{array}
 \end{equation}
where $\vartheta^{(k)}$ is defined in Proposition~\ref{proposition:relation_Vqk_Vq}.
Let $q\in\qdef$ and $k\in\Zge$. We have, in view of Proposition \ref{proposition:relation_Vqk_Vq} and \eqref{Eqt:proof_1:bellman_2},
\begin{equation}\label{equation_1:proof:corollary_3}
    \vartheta^{(1)}(Y(q),\tau+1)=Y(q)-\ualpha_{\tau}\circ\oalpha_{\tau}^{-1}(Y(q)).
\end{equation}
By items (ii) and (iii) of Corollary~\ref{corollary:expo:ell_discount} and since $\ell_2(\Zge)\subset\R_{>0}$, we have $\ualpha_{s_2}(s_1)=\uw_{s_2}(s_1)\geq\ual s_1\ell_2(s_2)$ and $\oalpha^{-1}_{s_2}(s_1)=(\ov_{s_2}+\ow_{s_2})^{-1}(s_1)\geq\tfrac{1}{\oav\ell_2(s_2)}s_1$ for any $s_1\geq0$ and $s_2\in\Zge$. Thus \eqref{equation_1:proof:corollary_3} implies 
\begin{align}\label{equation_2:proof:corollary_3}
    \vartheta^{(1)}(Y(q),\tau+1)&\leq Y(q)-\dfrac{\ual}{\oav\ell_2(\tau)}Y(q)\ell_2(\tau)\nonumber\\
    & = \Big(1-\dfrac{\ual}{\oav}\Big)Y(q).
\end{align}
We now show that $\ual\leq\oav$. By item (i) of Proposition~\ref{proposition:relation_Vqk_Vq}, we have $\uw\leq\ov+\ow$. Therefore, in view of items (ii) and (iii) of Corollary~\ref{corollary:expo:ell_discount}, $\ual s_1 \ell_2(s_2)\leq \oav s_1 \ell_2(s_2)$ for any $s_1\geq0$ and $s_2\in\Zge$, and thus $\ual\leq\oav$.
Therefore, \eqref{Equation_0:proof:corollary_3} is derived from  \eqref{equation_2:proof:corollary_3} by induction.

With similar steps as in the proof of Proposition~\ref{proposition:cost_gamma} and by taking $\oa(s_1)=\oav s_1$ and $\ua(s_1)=\ual s_1$ for any $s_1\geq 0$ and since $\oav\geq\ual$, we derive that 
$\ualpha_{k+\tau}^{-1}\left(\vartheta^{(k)}(Y(q),k+\tau)\right)\leq \tfrac{c_2\ogamma^{\tau}}{\ual c_1 \ugamma^{\tau}}\big(\tfrac{1}{\ugamma}\big(1-\tfrac{\ual}{\oav}\big)\big)^{k}\oav \sx$.
Therefore, item (ii) of Theorem~\ref{Theorem:stability} holds with, for any $s_1\geq 0$ and $s_2,s_3\in\Zge$,
     $\beta(s_1,s_2,s_3)=\lambda_1s_1\lambda_2^{s_2}\lambda_3^{s_3}$,
where $\lambda_1=\tfrac{\oav c_2}{\ual c_1}$, $\lambda_2=\tfrac{1}{\ugamma}\big(1-\tfrac{\ual}{\oav}\big)$ and $\lambda_3=\tfrac{\ogamma}{\ugamma}$, which is of class  $\ekl$ for any $s_3\in\Zge$. 
\end{proofa}

Note that Corollary~\ref{corollary:expo:ell_discount} does not require item (iv) of Proposition~\ref{proposition:cost_gamma} as it is ensured by items (ii) and (iii) of Corollary~\ref{corollary:expo:ell_discount}. 

\section{Examples}\label{Section:illustrative:example}
We present two examples to illustrate the results of Section~\ref{Section:ensuring_item_ii_theorem_1}. The first example presents a time-independent system with a time-varying cost, in which we compare our result with \cite[Example 3]{Mathieu_1_2018}. To highlight our contribution, the second example  considers a time-varying system and cost on which the previous results in the literature can not be applied.
\subsection{Non-holonomic integrator with time-varying cost}
Consider the time-invariant non-holonomic integrator as in \cite[Example 2]{grimm} 
\begin{flalign}
x_1(k+1)&=x_1(k)+u_1(k) \nonumber\\
x_2(k+1)&=x_2(k)+u_2(k) \\
x_3(k+1)&=x_3(k)+x_1(k)u_2(k)-x_2(k)u_1(k) \nonumber,
\end{flalign}
where $x=(x_1,x_2,x_3)\in\R^{3}$, $u=(u_1,u_2)\in\mathcal{U}(x)=\R^{2}$ and $k\in\Zge$. The stage cost is defined as $\ell(x,u,\tau)=(x_1^2+x_2^2+10|x_3|+|u|^2)\ell_2(\tau)$ and $\sx=x_1^2+x_2^2+10|x_3|$ for $x\in\R^3$, $\tau\in\Zge$ and $u\in\R^2$. 

\noindent\emph{Uniform global exponential stability.} Suppose that  $c\leq\ell_2(k)\leq d$ for any $k\in\Zge$ with $c,d>0$. We show that conditions of Corollary~\ref{corollary:expo:ell_lower_bounded} hold for this case. Indeed,
SA holds in view of \cite[Section~\RNum{4}-D]{Post_17}. Using the sequence of inputs constructed in \cite[Section~\RNum{4}-D]{Post_17}, we conclude that Condition~\ref{Assumption:Detectability} is verified with $W=0$, $\ow=0$ and $\uw(s_1,s_2)=s_1\ell_2(s_2)$ for any $s_1\geq0$ and $s_2\in\Zge$.  Condition~\ref{Assumption:Stabilizability}
is satisfied with $\ov(s_1,s_2)=\tfrac{22}{5}s_1 \ell_2(s_2)$  for any $s_1\geq0$ and $s_2\in\Zge$, see \cite[Section~\RNum{4}-D]{Post_17}. Item (i) of Corollary~\ref{corollary:expo:ell_lower_bounded} holds with $\ual=1$ and $\uw(s_1,s_2)=s_1\ell_2(s_2)$ for any $s_1\geq0$ and $s_2\in\Zge$. Item (ii) of the same corollary holds with $\oav=\tfrac{22}{5}$, $\ow=0$ and $\ov(s_1,s_2)=\tfrac{22}{5}s_1 d$  for any $s_1\geq0$ and $s_2\in\Zge$. Therefore, conditions of Corollary~\ref{corollary:expo:ell_lower_bounded} are verified and \eqref{equation:theorem:KL_stability:item_2}  applies with  $\beta$ independent of $\tau$ and $\beta\in\ekl$.

\noindent\emph{When $\ell(q,u)=\ell_1(x,u)\ell_2(\tau)$ for any $q=(x,\tau)\in\mathcal{Q}$ and $u\in\mathcal{U}(q)$.} Suppose that $\ugamma^k\leq\ell_2(k)\leq\ogamma^k$ for any $\mathit{k\in\Zge}$ with $\ogamma,\ugamma>0$. We show that conditions of Corollary~\ref{corollary:expo:ell_discount} hold for this case. Item (i) of Corollary~\ref{corollary:expo:ell_discount}  is verified in view of the expression of $\ell$. In view of the above developments, we have  items (ii) and (iii) of Corollary~\ref{corollary:expo:ell_discount} are satisfied with $\ua_{\ell}=1$ and $\oa_V=\tfrac{22}{5}$, respectively. Furthermore, item (iv) of Corollary~\ref{corollary:expo:ell_discount} is satisfied if $\ugamma,\ogamma\in(1-\tfrac{5}{22},\infty)$ and $c_1=c_2=1$. Therefore, all the items of Corollary~\ref{corollary:expo:ell_discount} are verified and thus \eqref{equation:theorem:KL_stability:item_2} holds with  $\beta(\cdot,\cdot,s)\in\exp-\mathcal{KL}$ for any $s\in\Zge$. For the case of the discounted cost where   $\ell_2(k)=\gamma^k$ with $\gamma\in(0,1)$ for any $k\in\Zge$, item (iv) of Corollary~\ref{corollary:expo:ell_discount} is satisfied with $\ogamma=\ugamma=\gamma$ and $c_1=c_2=1$. We find that \eqref{equation:theorem:KL_stability:item_2} holds with $\beta\in\exp-\mathcal{KL}$ if $\gamma\in\big(\tfrac{17}{22},1\big)$. This bound corresponds to the one obtained in \cite[Example 3]{Mathieu_1_2018},  in which discounted costs are studied.

\subsection{Tracking control of an inverted pendulum}
In this example, we investigate the optimal control of an inverted pendulum where the objective is
to optimally track a given reference trajectory, which is a classical problem in control theory and robotics.
We consider the model of an inverted pendulum discretized
by an Euler scheme with sampling period $T>0$,
\begin{flalign}
\hspace{1cm}z_1(k+1)&=z_1(k)+Tz_2(k),\\
z_2(k+1)&=z_2(k)+T(a\sin(z_1(k))-bz_2(k)+cu(k)),\nonumber
\end{flalign}
where $z_1(k)\in\R$ is the angular position of the pendulum, with
$z_1 = 0$ being the upper-position, $z_2(k)\in\R$ is the angular velocity
and $u(k)\in\R$ is a controllable torque at the rotation axis at time $k\in\Zge$. The
constants $a$, $b$, $c>0$ are related to the mass, the dissipation
and the motor gain, respectively. We define $z:=(z_1,z_2)$. 
The objective is to optimally track a reference trajectory $z_{\text{ref}}=(z_{\text{ref}_1},z_{\text{ref}_2})$ satisfying 
   $z_{\text{ref}_1}(k+1)=z_{\text{ref}_1}(k)+Tz_{\text{ref}_2}(k),$ 
    $z_{\text{ref}_2}(k+1)=z_{\text{ref}_2}(k)+T(a\sin(z_{\text{ref}_1}(k))-bz_{\text{ref}_2}(k)+cv(k)),$
where $v$ is the reference input taking values in a bounded set. We define the tracking error $e(k):=z(k)-z_{\text{ref}}(k)$, whose dynamics is
\begin{flalign}
  \hspace{1cm}  e_1(k+1)=&e_1(k)+Te_2(k),\\
    e_2(k+1)=&e_2(k)+T(a\sin(e_1(k)+z_{\text{ref}_1}(k))\nonumber\\
            &-a\sin(z_{\text{ref}_1}(k))-be_2(k)+c(u(k)-v(k))).\nonumber
\end{flalign}

Consider the state $x:=(e,z_{\text{ref}})$. Let $\sigma(x) = |e_1| + |e_2|$ and the stage cost $\ell(x, u,\tau) = \uell(\tau)(\sigma(x) + r|u-v(\tau)|)$ with $r>0$ and $\uell(\tau)\in[\underline{m},\overline{m}]$ with $0<\underline{m}\leq \overline{m}<\infty$ for any $x\in \R^4$, $\tau\in\Zge$ and $u\in\R$. First, we verify that SA holds by applying  \cite[Theorem 1 and Theorem 2(\text{$d_3$})]{gilbert}. We have that items a)-c) of Theorem 1 in \cite{gilbert} are verified. Item $d_3$) of Theorem 2 in \cite{gilbert}
is satisfied  by considering $\phi_k(u)=r|u-v(\tau)|$ with the notation of \cite{gilbert}, as $v$ takes value in a bounded set. For item e) of Theorem 1 in \cite{gilbert}, let $x\in\R^4$
and consider the infinite sequence $\textbf{u}=\Big(v(0)+\tfrac{1}{c}\big(be_2-a(\sin(e_1+z_{\text{ref}_1})-\sin(z_{\text{ref}_1}))-\tfrac{1}{T}e_2-\tfrac{e_1+Te_2}{T^2}\big), v(1)+\tfrac{1}{c}\big(-b\big(\tfrac{e_1+Te_2}{T}\big)-a(\sin(e_1+Te_2+z_{\text{ref}_1}+Tz_{\text{ref}_2})-\sin(z_{\text{ref}_1}+Tz_{\text{ref}_2}))+\tfrac{e_1+Te_2}{T^2}\big),v(2),v(3),\cdots\Big)$. It follows that $\phi(0,x,\textbf{u}|_1)=(e_1+Te_2,-\tfrac{e_1+Te_2}{T},z_{\text{ref}_1}+Tz_{\text{ref}_2},z_{\text{ref}_2}+T(a\sin(z_{\text{ref}_1}-bz_{\text{ref}_2}+cv(0)))$ and $\phi(k,x,\textbf{u}|_k)=(0,0,z_{\text{ref}_1}(k),z_{\text{ref}_2}(k))$ for any $k\geq1$. We deduce that the cost \eqref{E:S:intial_cost} is finite, hence item e) of Theorem 1 in \cite{gilbert} applies. We can then apply \cite[Theorems 1 and 2]{gilbert} to deduce that SA holds. We now investigate Condition~\ref{Assumption:Stabilizability}. Let $q=(x,\tau)\in\R^4\times\Zge$, and consider the same sequence of inputs $\textbf{u}$ as above. We derive that $J(q,\textbf{u}) \leq \theta \sigma(x)$ with $\theta>0$ independent of $q$ and $\textbf{u}$. Hence, Condition~\ref{Assumption:Stabilizability} is satisfied
with $\ov(s_1,s_2) = \theta s_1$ for any $s_1\geq0$ and $s_2\in\Zge$. From $\ell(x, u,v) \geq \underline{m}\sx$, Condition~\ref{Assumption:Detectability} is verified with $W = \ow = 0$ and $\uw(s_1,s_2) = \underline{m} s_1$ for any $s_1\geq0$ and $s_2\in\Zge$.

Now, we show that conditions of Corollary~\ref{corollary:expo:ell_lower_bounded} hold. Item (i) of Corollary~\ref{corollary:expo:ell_lower_bounded} holds with $\ual=\underline{m}$ and $\uw(s_1,s_2)=\underline{m}s_1$ for any $s_1\geq0$ and $s_2\in\Zge$. Item (ii) of the same corollary holds with $\oav=\max\{\theta_1(r,T),\theta_2(r,T)\}$, $\ow=0$ and $\ov(s_1,s_2)=\max\{\theta_1(r,T),\theta_2(r,T)\}s_1$  for any $s_1\geq0$ and $s_2\in\Zge$. Therefore, conditions of Corollary~\ref{corollary:expo:ell_lower_bounded} are verified and \eqref{equation:theorem:KL_stability:item_2}  applies with  $\beta\in\mathcal{\ekl}$.

\section{Conclusion}\label{section:conculsion}
We have analyzed the stability of deterministic time-varying nonlinear discrete-time systems for which the sequence of inputs aim to minimize an infinite-horizon time-dependent cost. We provided conditions under which (non-)uniform asymptotic and exponential stability properties are guaranteed for this class of systems. These results generalize the recent contributions on the stability analysis of discounted optimal control problems \cite{Gaitsgory_global_stability_ct,Post_17} to more general time-dependent costs. An interesting future work would be to extend these results to the stochastic setting. 
\bibliographystyle{plain}        
\bibliography{autosam}  
\end{document}